\documentclass[conference]{IEEEtran}
\IEEEoverridecommandlockouts
\usepackage{cite}
\usepackage{amsmath,amssymb,amsfonts}
\usepackage{algorithmic}
\usepackage{graphicx}
\usepackage{textcomp}
\usepackage{xcolor}
\usepackage{booktabs}
\usepackage{multirow}
\def\BibTeX{{\rm B\kern-.05em{\sc i\kern-.025em b}\kern-.08em
    T\kern-.1667em\lower.7ex\hbox{E}\kern-.125emX}}
\begin{document}

\title{USTC-NELSLIP System Description for DIHARD-III Challenge\\
}

\author{
\IEEEauthorblockN{Yuxuan Wang}
\IEEEauthorblockA{
\textit{University of Science and Technology}\\
\textit{of China}\\
Hefei, Anhui, China \\
yxwang1@mail.ustc.edu.cn}
\and
\IEEEauthorblockN{Maokui He}
\IEEEauthorblockA{
	\textit{University of Science and Technology}\\
	\textit{of China}\\
	Hefei, Anhui, China \\
	hmk1754@mail.ustc.edu.cn}
\and
\IEEEauthorblockN{Shutong Niu}
\IEEEauthorblockA{
	\textit{University of Science and Technology}\\
	\textit{of China}\\
	Hefei, Anhui, China \\
	niust@mail.ustc.edu.cn}
\and
\IEEEauthorblockN{Lei Sun}
\IEEEauthorblockA{
	\textit{iFlytek Research}\\
	Hefei, Anhui, China \\
	leisun8@iflytek.com}
\and
\IEEEauthorblockN{Tian Gao}
\IEEEauthorblockA{
	\textit{iFlytek Research}\\
	Hefei, Anhui, China \\
	tiangao5@iflytek.com}
\and
\IEEEauthorblockN{Xin Fang}
\IEEEauthorblockA{
	\textit{iFlytek Research}\\
	Hefei, Anhui, China \\
	xinfang@iflytek.com}

\and
\IEEEauthorblockN{Jia Pan}
\IEEEauthorblockA{
	\textit{iFlytek Research}\\
	Hefei, Anhui, China \\
	jiapan@iflytek.com}
\and
\IEEEauthorblockN{\qquad\qquad\qquad\qquad Jun Du}
\IEEEauthorblockA{
	\textit{\qquad\qquad\qquad\qquad University of Science and Technology of China}\\
	\qquad\qquad\qquad\qquad Hefei, Anhui, China \\
	\qquad\qquad\qquad\qquad jundu@mail.ustc.edu.cn}
\and
\IEEEauthorblockN{Chin-Hui Lee}
\IEEEauthorblockA{
	\textit{Georgia Institute of Technology}\\
	Atlanta, Georgia, USA \\
	chl@ece.gatech.edu}
}
\maketitle 

\begin{abstract}
This system description describes our submission system to the Third DIHARD Speech Diarization Challenge. 
Besides the traditional clustering based system, the innovation of our system lies in the combination of various front-end techniques to solve the diarization problem, including speech separation and target-speaker based voice activity detection (TS-VAD), combined with iterative data purification.
We also adopted audio domain classification to design domain-dependent processing. Finally, we performed post processing to do system fusion and selection. Our best system achieved DERs of 11.30\% in track 1 and 16.78\% in track 2 on evaluation set, respectively.
\end{abstract}


\section{Notable Highlights}

1. Iterative speech separation (ISS) based diarization system.

2. Iterative TS-VAD (ITS-VAD) based diarization system.

3. Domain-dependent processing.

\section{Data Resources}

The overall framework of our system is shown in Figure 1. In this section, we introduce all data sources for submodules:

\textbf{Audio Domain Classification:} The whole DIHARD-III development set (LDC2020E12) was divided into 2 parts, 9/10 for training and 1/10 for testing.

\textbf{Speech Enhancement:} The clean speech data were from WSJ0\cite{paul1992design}, AIShell-1\cite{bu2017aishell}, THCHS-30\cite{wang2015thchs}, and Librispeech\cite{panayotov2015librispeech}, and the noise data included 115 types of noise\cite{sun2018novel} and MUSAN\cite{snyder2015musan} corpus. The input noisy mixture was made at -5 dB, 0 dB and 5 dB, and finally we got a 1000-hour training set.

\textbf{Clustering Based Diarization System:} We directly referred to the system of BUT in\cite{landini2020but}. For SAD, we used oracle SAD results for track 1, and scratched SAD results for track 2. The pre-trained model was trained on the 600-hour home-made realistic speech data in iFlytek. The speech quality is not very stable due to the complicated acoustic environments. And the DIHARD-III development set was used for finetuning. For x-vector extraction, the training data was drawn from VoxCeleb 1\cite{nagrani2017voxceleb} and 2\cite{chung2018voxceleb2} with data augmentation, amounting to 6 million utterances from 7146 speakers. For PLDA, the out-of-domain PLDA was trained with VoxCeleb 1 and 2, and the in-domain PLDA using DIHARD-II development set (LDC2019E31). 
\begin{figure}[h]
	\centering
	\includegraphics[width=\linewidth]{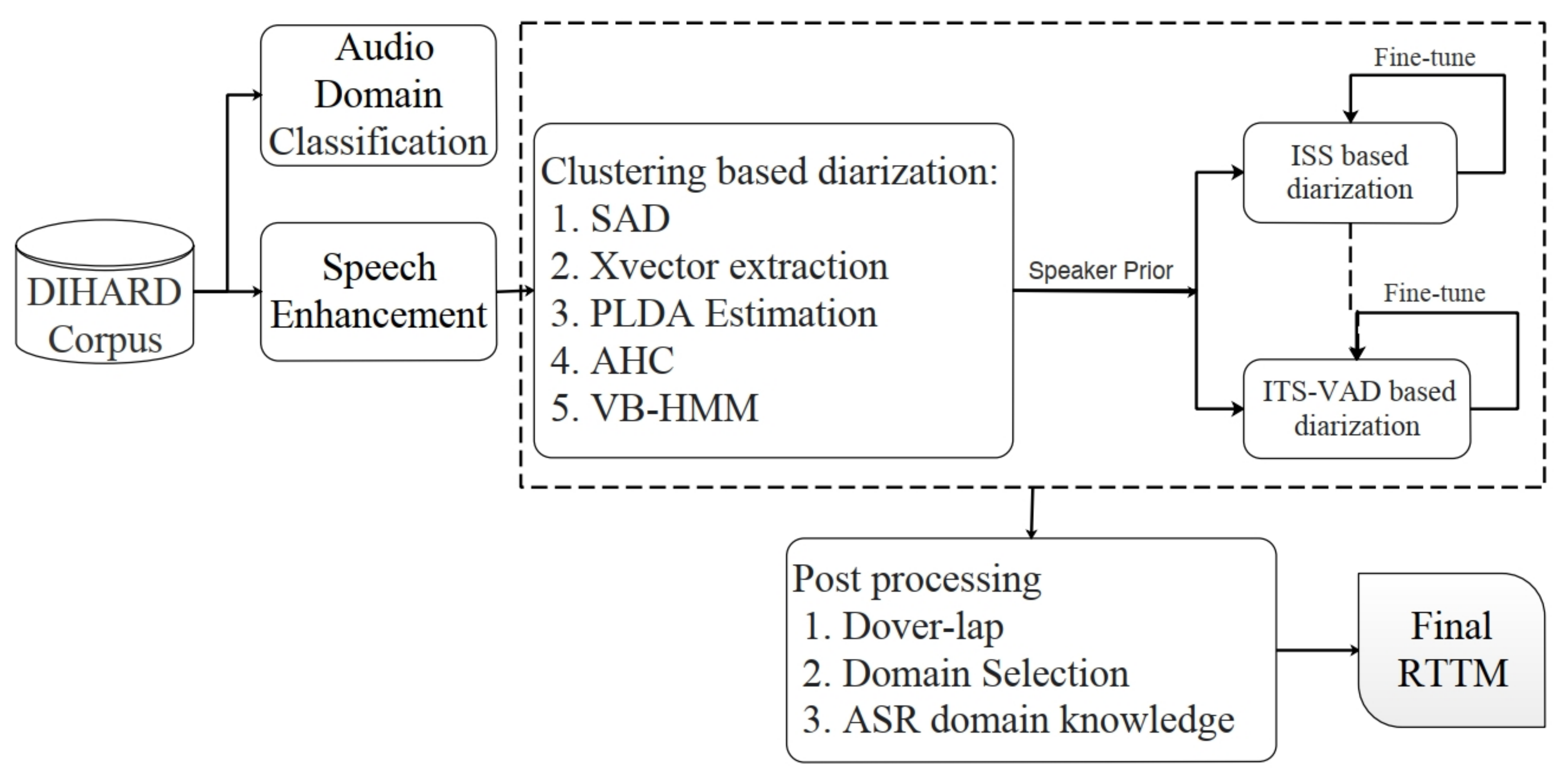}
	\caption{An illustration of overall framework.}
	\label{fig:system_overview}
\end{figure}

\textbf{ISS Based Diarization System:} The pre-trained model was trained on the 250-hour 2-speaker mixed data simulated by Librispeech.

\textbf{ITS-VAD Based Diarization System:} We used VoxCeleb 1 and 2 with augmented with 3-fold speed perturbation to train the i-vector extractor. The pre-trained model was trained on the 2500-hour data, in which Switchboard-2 (LDC97S62), AMI Meeting Corpus\cite{mccowan2005ami}, Voxconverse\cite{chung2020spot} DEV set were selected as realistic conversations and Librispeech as simulated meeting-style multi-speaker conversations.

\textbf{ASR:} The whole CHIME-6 training set\cite{watanabe2020chime} was used for ASR model training.


\section{Detailed Description of Algorithm}

\subsection{Audio Domain Classification}\label{AA}
We used a ResNet based network with 17 convolutional layers\cite{he2016deep}\cite{mcdonnell2020acoustic} to classify the diverse 11 domains both in the development set and evaluation set. 64-dimensional log-mel filterbanks were used as input acoustic features, and 11 output-layer cells corresponded to 11 domains. During training, all sessions were truncated into 10-second segments, each segment was assgined with the corresponding domain label, the initial learning rate was set to 0.1, and the batch-size was set to 15. We trained it for 510 epochs.  During testing, we used voting strategy to get session-level classifciation results. The proposed model achieved 100\% accuracy on the development set, and we directly applied it on the evaluation set to predict the domain for each session. 

\subsection{Speech Enhancement}
We employed the progressive multi-target network based speech enhancement model\cite{sun2020progressive} on RESTAURANT domain for its loud background noise. 257-dimensional PELPS and PRM were selected simultaneously.
 7-frame expansion was used for input, the number of LSTM memory cells in each layer was 1024, and the progressive increasing SNR between two adjacent targets was set to 10 dB. Besides, we applied speech enhancement on all domains before feeding into SAD module in Track2. Here $PELPS_{1}$ (Progressively Enhanced LPS at target layer 1) enhanced speech was found optimal for the following diarization results and SAD results.

\subsection{Clustering Based Diarization System}

The traditional clustering based diarization system was used as the basis system for the whole framework, and the diarization 
results were used as initialization for the subsequent submodules except the RESTAURANT domain. Experiment results showed that performing SS or TS-VAD based diarization on RESTAURANT domain data led to the performance degradation, so we collected the results of clustering based diarization system of RESTAURANT domain as the final submitted results directly. 

The parameters finetuned on development set for RESTAURANT domain were set as followings: the interpolation value of the two PLDAs was 0.57, that is, the combination weight of the in-domain PLDA was 0.57, and the out-of-domain PLDA 0.43, the thresholds used as stopping criteria for the AHC were finetuned as $threshold\_bias$ = 0.5, $target\_energy$ = 0.3, the parameters for the BHMM was finetuned as $maxIters$ = 7, $smoothing\_factor$ = 4.0, $lda\_dim$ = 512. 
The rest parameters on RESTAURANT domain and all the parameters on other domains kept consistent with \cite{landini2020but}. 

For SAD, We employed three different networks for framewise binary classification of speech and non-speech. The DNN model adopted a small and compact structure using 2 hidden layers with 256 and 128 hidden units in each layer and a final dual output layer. The input features were 39-dimensional PLP features and expanded an input context of 5 neighbouring frames (±2). The CLDNN model\cite{sainath2015convolutional} adopted 2 CNN layers, 2 LSTM layers and 2 DNN layers. The input features were 40-dimensional filterbanks. The TDNN model was equipped with the same structure as the DIHARD-III SAD baseline\cite{ryant2020third}. As mentioned in III-B, all of those models were finetuned on enhanced DIHARD-III development set and tested on enhanced DIHARD-III evaluation set (LDC2020E13).
Finally, we voted from the three systems and got a fusion SAD result. 
\subsection{ISS Based Diarization System}

The SS framework simply contains two parts: separation and detection. 
In the separation part, we trained a fully convolutional time-domain audio separation network (Conv-TasNet)\cite{luo2019conv} as our pre-trained model. 
In the detection part,  we directly used a DNN-based SAD to detect speaker presence.  Combine all SAD results along the time axis, then speech diarization results were attained, and the overlap regions were automatically labeled.
\begin{figure}[h]
	\centering
	\includegraphics[width=0.5\textwidth]{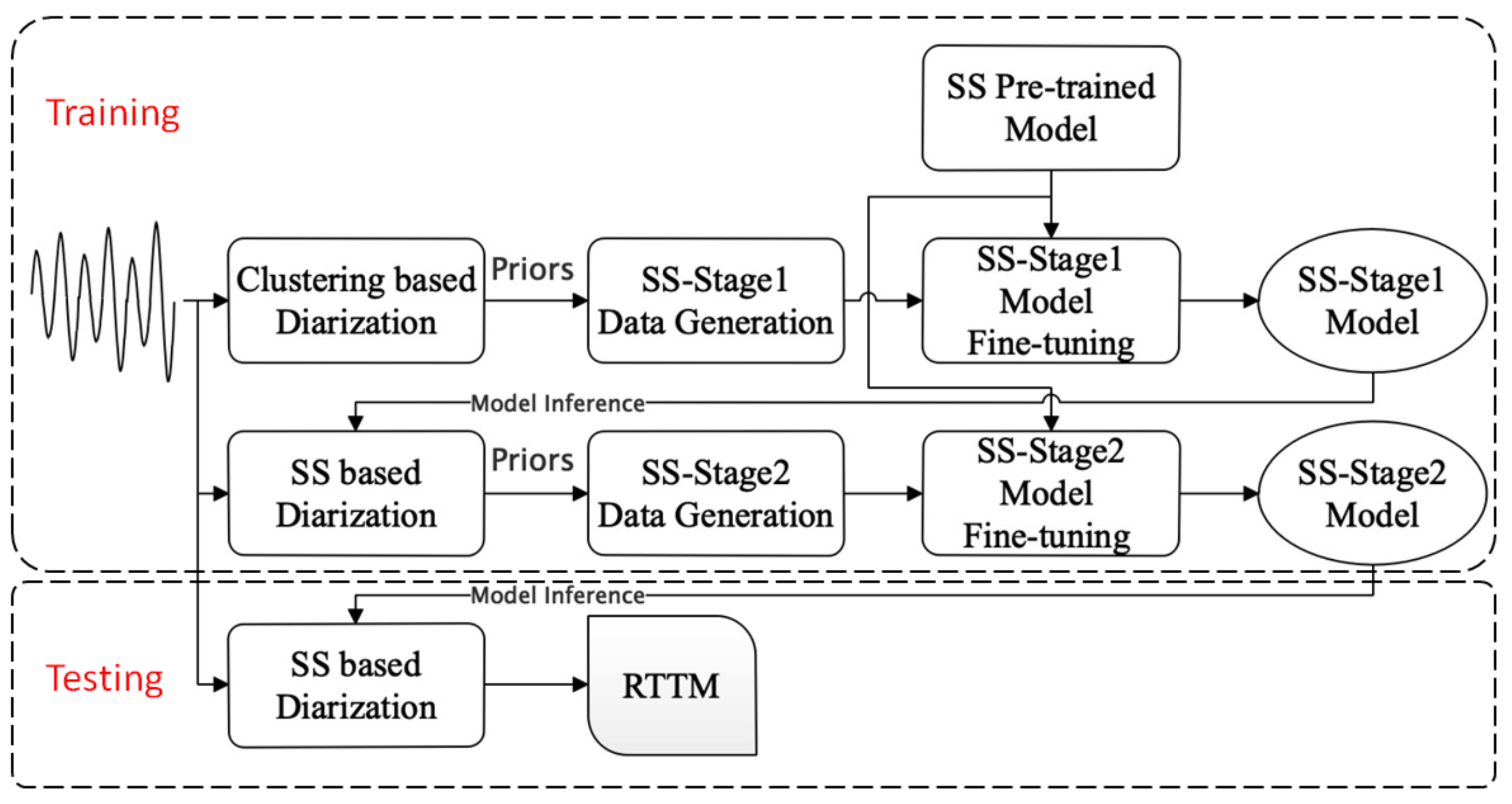}
	\caption{The training and testing procedure for ISS Based Diarization System}
	\label{fig:spectragram}
\end{figure}

To improve the generalization ability of the separation model, we adopted the iterative strategy, that is, we used the speaker priors drawn from the traditional clustering based diarization system to simulate data, about 5000 mixed audios for each session, and retrained to let the SS pre-trained model adapt current session. Then we got the SS-stage1 model. 
The same procedure was taken once more except the speaker priors were drawn from SS-stage1 model, which can significantly improve the separation performance by the more accurate priors. Finally, we used the SS-stage2 model to separate the entire session. To be mentioned that we finetuned the separation model for each session. The whole framework is shown in figure 2.
Here we only applied this method on CTS domain in our submitted system.

The asteroid\cite{Pariente2020Asteroid} was used as our speech separation toolkit to train the Conv-TasNet model. In the pre-training process, the learning rate was set to 0.001, and the batch-size was set to 6. We trained it for 75 epochs. In the finetuning process, the learning rate was also set to 0.001, the batchsize was set to 4 and we trained it for 3 epochs.

\subsection{ITS-VAD Based Diarization System}

The TS-VAD model\cite{medennikov2020target} takes log mel filterbanks as input, along with i-vectors corresponding to the speakers, and predicts per-frame speech activities for all the speakers simultaneously. 

When the speaker number in a session was smaller than the number of output nodes $N$, we assigned the remain nodes to dummy speakers selected from the training set randomly, labeled them with silence when training and discarded them when testing. When the number of speakers was larger than $N$,  we randomly selected $N$ speakers for training and selected $N$ speaker with longest voice for testing. All the speaker numbers above were estimated from traditional clsutering based diarization systems, and we set $N$ to 8 here.

To improve the generalization ability of the TS-VAD model, we also applied iterative strategy here. First, we decoded the TS-VAD pre-trained model with i-vectors extracted from the clustering based diarization results. Then, we removed overlapping segments detected by TS-VAD from clustering based results and used the remaining non-overlapping segments to simulate multiple speaker dialogue data, about 4 hours for each session, for TS-VAD finetuning. Finally, we used the i-vectors extracted from the non-overlapping segments to decode the TS-VAD finetuned model and generated the iterative TS-VAD diarization results. Also we finetuned TS-VAD pre-trained model for each session. The whole framework is shown in figure 3. The number of iterations for iterative training depended on the different domains.

We implemented TS-VAD related experiments with kaldi and pytorch toolkit. In pre-training process, the training data was divided into 8 seconds each segment, the learning rate of the adam optimizer was set to 0.0001, and the batch-size was set to 32. And we trained it for 2 epochs. In finetuning process, the learning rate was also set to 0.0001, and the batch-size was set to 8.
\begin{figure}[h]
	\centering
	\includegraphics[width=0.5\textwidth]{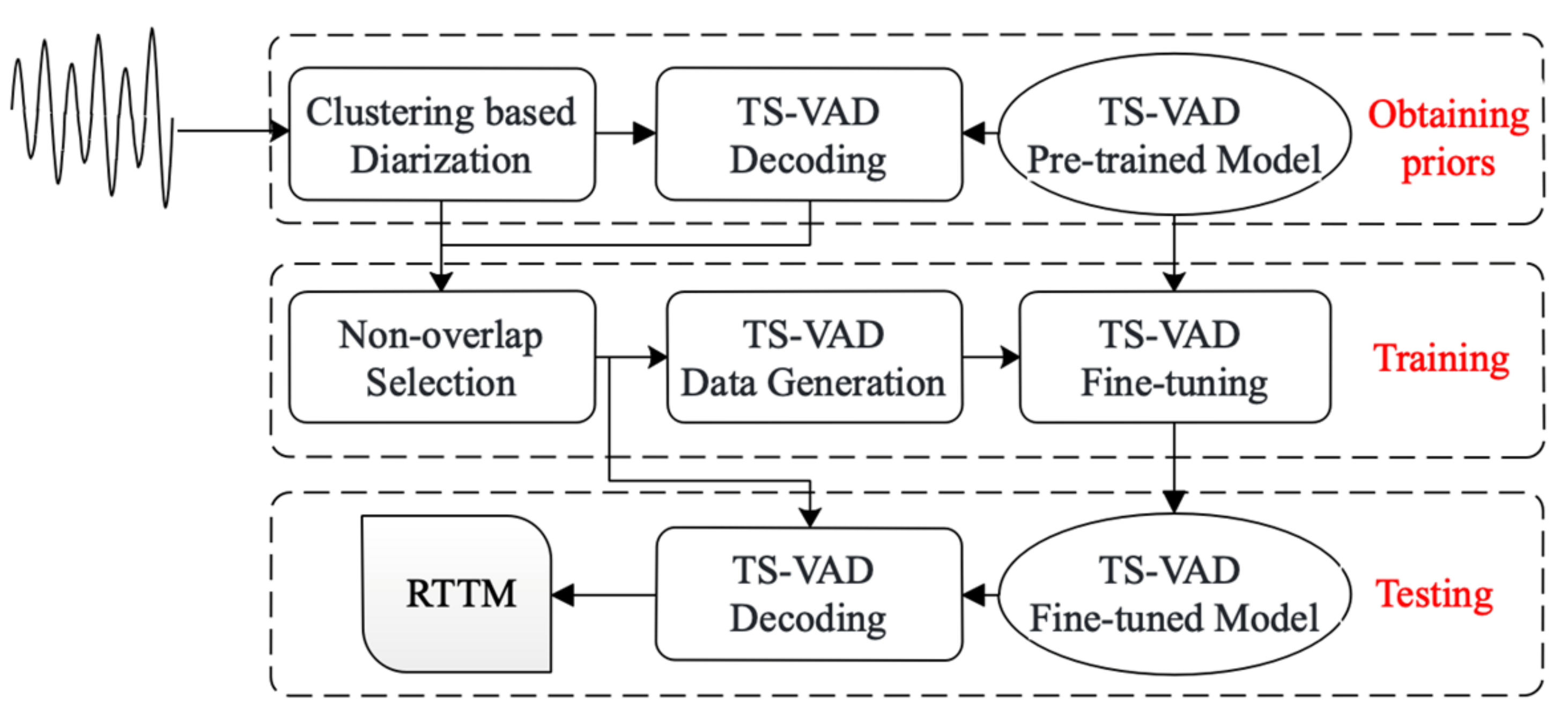}
	\caption{The training and testing procedure for ITS-VAD Based Diarization System}
	\label{fig:spectragram}
\end{figure}
\subsection{Post-processing}

 As mentioned before, we got several diarziation systems including clustering based diarization, ISS based diarization and ITS-VAD based diarization with different priors. We first performed system fusion with Dover-lap\cite{raj2020dover} of above results, which can effectively improve overall performance. Dover-lap here was not only used for fusing different systems, but also fusing different epochs in the same system during the iterative processing.
 
 Based on  the results of audio domain classification, we also performed domain selection, that is, select the best system and the best set of parameters for each domain according to the performance on the development set.
 
Specially, we utilized domain knowledge from the speech recognition task,   the recognized token [laugh] indicates where overlapping speech segments often occur, especially in multi-talker scenarios. We simply assigned laughter segments to each neighborhood speaker. This also gave a slight performance improvement.

\section{Results}
In this section, we introduce the performance of the submission systems and other major system components.
\begin{table}[h]
	\centering
	\caption{Performance comparision on full development set CTS data for TRACK 1 among different systems.}
	\setlength{\tabcolsep}{1mm}{
	\begin{tabular}{ccccc}
	\toprule
	System    &Miss(\%) &FA(\%) &SpkErr(\%)             & DER(\%) \\ \hline
	Clustering based diarization system&12.00 &0.00&4.22 & 16.22       \\ \hline
	\begin{tabular}[c]{@{}c@{}}Clustering based diarization system\\ +SS based diarization system\end{tabular}  &7.61&2.61&2.73            & 12.95 \\ \hline
	\begin{tabular}[c]{@{}c@{}}Clustering based diarization system\\ +ISS based diarization system\end{tabular} &5.12&1.74&1.90             & 8.76 \\ \hline
	\begin{tabular}[c]{@{}c@{}}Clustering based diarization system\\ +ISS based diarization system\\ +Dover-lap\end{tabular}&5.40&1.21&1.70 & 8.31 \\ \bottomrule
	\end{tabular}}
\end{table}
\begin{table*}[t]
	\centering
	\caption{DER(\%) Performance comparision on domain-wise full development set for TRACK 1 among different systems.}
	\resizebox{\linewidth}{!}{
		\begin{tabular}{ccccccccc}
			\toprule
			System &
			{\color[HTML]{000000} MAPTASK} &
			{\color[HTML]{000000} BROADC.} &
			{\color[HTML]{000000} COURT.} &
			{\color[HTML]{000000} SOC. LAB} &
			{\color[HTML]{000000} CTS} &
			{\color[HTML]{000000} CLINICAL} &
			{\color[HTML]{000000} SOC. FIELD} &
			{\color[HTML]{000000} MEETING}  \\ \hline
			Clustering based diarization system &
			{\color[HTML]{000000} 5.02} &
			{\color[HTML]{000000} 2.60} &
			{\color[HTML]{000000} 2.95} &
			{\color[HTML]{000000} 7.97} &
			{\color[HTML]{000000} 16.22} &
			{\color[HTML]{000000} 10.97} &
			{\color[HTML]{000000} 11.87} &
			{\color[HTML]{000000} 26.41}  \\ \hline
			\multicolumn{1}{l}{\begin{tabular}[c]{@{}l@{}}Clustering based diarization system\\ +TS-AVD based diarization system\end{tabular}} &
			{\color[HTML]{000000} 6.71} &
			{\color[HTML]{000000} 2.94} &
			{\color[HTML]{000000} 3.15} &
			{\color[HTML]{000000} 8.81} &
			{\color[HTML]{000000} 10.21} &
			{\color[HTML]{000000} 16.48} &
			{\color[HTML]{000000} 13.79} &
			{\color[HTML]{000000} 24.72}  \\ \hline
			\begin{tabular}[c]{@{}c@{}}Clustering based diarization system\\ +ITS-VAD based diarization system\end{tabular} &
			{\color[HTML]{000000} 2.27} &
			{\color[HTML]{000000} 2.37} &
			{\color[HTML]{000000} 2.46} &
			{\color[HTML]{000000} 5.17} &
			{\color[HTML]{000000} 7.76} &
			{\color[HTML]{000000} 9.83} &
			{\color[HTML]{000000} 10.74} &
			{\color[HTML]{000000} 23.05} \\  \bottomrule
	\end{tabular}}
\end{table*}

Table I lists the performance comparison of different systems on CTS domain of full development set with oracle SAD. CTS data is collected in telephone environment, and has 2 speakers in a session with quite a bit of overlapping segments. We get a DER of 16.22\% for the first clustering based diarization system, and the miss value is observed to be the largest part in the whole DER, which indicates that the traditional algorithm cannot well handle overlapped speech.
To solve this problem, we add the second SS based diarization system (pre-trained model), the miss value achieves a large decrease of 4.39\% (from 12.00\% to 7.61\%), and the increased FA value is due to the separation error, which may create additional sounds on the original speech segment. The corresponding DER also decreases from 16.22\% to 12.95\%. Next, when we add the third ISS based diarization system (SS-stage2 model), the performance is further improved, the DER changes from 12.95\% to 8.76\% with the decrease of 4.19\%. The last system we add dover-lap, the dover-lap here is used to fuse different iterations during iterative training, and we finally get a DER of 8.31\%, totally 48.7\% relative improvement over the first traditional clustering based system.

Table II compares the DER performance on domain-wise full development set for track 1 except three domains: AUDIOBOOKS, WEBVIDEO, and RESTAURANT. There is only one speaker for each session in AUDIOBOOKS, so we simply apply SAD processing, and assign each speech segment with the same speaker in one session. For RESTAURANT domain, as mentioned above, it has many speakers and overlapped speech with loud background noise, TS-VAD leads to performance degradation. So does WEBVIDEO domain. From the results we can learn that, TS-VAD based diarization system (pre-trained model) only performs better than clustering based diarzation on well matched domains like CTS and MEETING (a decrease of 6.01\% and 1.69\%, respectively) corresponding to Switchboard and AMI in training set, respectively. While when we perform ITS-VAD based diarization system here, considerable improvement is brought on all the eight domains, which shows its huge generalization abilities. Note that the iterative training adopts the priors from ISS based diarization system instead of clustering based one for CTS domain here.
\begin{table}[h]
	\centering
	\caption{THE OVERALL ERROR (\%) of different SAD systems for the core/full DEV and EVAL
		sets. The Part. columns indicates whether scoring was performed using the full or core DEV/EVAL set.}
	\setlength{\tabcolsep}{5mm}{
		\begin{tabular}{cccc}
			\toprule
			\multicolumn{1}{l}{system} & part. & dev & eval \\ \hline
			\multirow{2}{*}{DNN}       & core  & 1.55    & \//     \\
			& full  & 1.51    &  6.18    \\ \hline
			\multirow{2}{*}{CLDNN}     & core  & 1.60    & \//     \\
			& full  & 1.64    &  6.55    \\ \hline
			\multirow{2}{*}{TDNN}      & core  & 1.44    & \//     \\
			& full  & 1.42    &  6.16    \\ \hline
			\multirow{2}{*}{Fusion}    & core  &1.32     & \//     \\
			& full  & 1.36    & 5.27    \\ \hline
			\multirow{2}{*}{Baseline}    & core  & 2.30    & 7.26     \\
			& full  & 2.42    & 6.51     \\ \bottomrule
	\end{tabular}}
\end{table}

Table III shows the overall error of the three single and the fusion SAD models. Different structures complement each other in the detection performance, so we use a weighted combine to fuse the single models, and achieve an error improvement of 1.06\% (from 2.42\% to 1.36\%) over baseline SAD on full development set. The same fusion strategy is applied on the full evaluation set, and we get an error improvement of 1.24\% (from 6.51\% to 5.27\%).
\begin{table}[h]
	\centering
	\caption{THE DER (\%) AND JER(\%) comparision of baseline and USTC-NELSLIP BEST SUBMISSION system for the core/full DEV and EVAL sets iN TRACK 1. }
	\setlength{\tabcolsep}{3mm}{
		\begin{tabular}{cccccc}
			\toprule
			\multirow{2}{*}{System} & \multirow{2}{*}{part.} & \multicolumn{2}{c}{DER(\%)} & \multicolumn{2}{c}{JER(\%)} \\
			&      & dev & eval & dev & eval \\ \hline
			\multirow{2}{*}{baseline} & core & 20.25    & 20.65     & 46.02    &  47.74    \\
			& full &19.41&19.25&41.66&42.45     \\ \hline
			\multirow{2}{*}{USTC-NELSLIP} & core & 13.30    & 13.45     & 34.05    &  34.94    \\
			& full & 11.07    & 11.30     &29.48     & 29.94     \\  \bottomrule
	\end{tabular}}
\end{table}
\begin{table}[!ht]
	\centering
	\caption{THE DER (\%) AND JER(\%) comparision of baseline and USTC-NELSLIP BEST SUBMISSION system for the core/full DEV and EVAL sets iN TRACK 2. }
	\setlength{\tabcolsep}{3mm}{
		\begin{tabular}{cccccc}
			\toprule
			\multirow{2}{*}{System} & \multirow{2}{*}{part.} & \multicolumn{2}{c}{DER(\%)} & \multicolumn{2}{c}{JER(\%)} \\
			&      & dev & eval & dev & eval \\ \hline
			\multirow{2}{*}{baseline} & core &22.28&27.34&47.75&51.91   \\
			& full&21.71&25.36&43.66&46.95     \\ \hline
			\multirow{2}{*}{USTC-NELSLIP} & core & 14.49    & 19.37     &  35.19   &  39.22    \\
			& full & 13.25   & 16.78     & 31.25    & 34.42     \\  \bottomrule
	\end{tabular}}
\end{table}

Table IV and V lists the performance of baseline and our best submission systems in track 1 and track 2, respectively. For both tracks, both sets and both metrics, our systems all significantly outperform baseline, and finally we ranked 1st on the leaderboard among all participants.
\section{Hardware Requirements}
The infrastructure used to run the experiment was a GPU, TeslaV100-PCIE, with a total memory of 12GB unless specified otherwise.

For audio domain classification, the processing time for the full development set is about 3 minutes, among which about 2 minutes to get the segment-level results, and 1 minute to vote to get senssion-level results and save. 

For speech enhancement, the processing time for 10 minutes of audio is about 20s, so the full development set cost about 200 minutes.

The clustering based diarization system processes the full development set using about 6 hours and 40 minutes, among which the time for AHC and BHMM clustering is about 4 hours and 20 minutes.

The ISS based diarization system processes 10 minutes of audio using about 20 seconds, and the total full development set CTS data using about 20 minutes.

The ITS-VAD based diarization system processes 10 minutes of audio using about 10 seconds.

The DNN-SAD, CLDNN-SAD, and TDNN-SAD model process the full development set using about 40 seconds, 45 seconds, and 45 seconds, respectively.

The post processing totally uses about 2 minutes to process the full development set.

\end{document}